\documentclass[letterpaper]{JHEP}

\usepackage{amssymb,amsmath}
\usepackage{epsfig}

\def\be{\begin{eqnarray}}
\def\ee{\end{eqnarray}}

\def\3{\ss}

\def\half {\frac{1}{2}}

\preprint{{\tt hep-th/0402124}}

\title{Imaginary time D-branes to all orders}

\author{Oren Bergman and Shlomo S.~Razamat\\
Department of Physics\\
Technion, Israel Institute of Technology\\
Haifa 32000, Israel\\
\email{bergman, razamat@physics.technion.ac.il}}

\abstract{Extending the work of Gaiotto, Itzhaki and Rastelli in
hep-th/0304192, we derive a general prescription for computing
amplitudes involving a periodic array of D-branes in imaginary time to
arbitrary order. We use this prescription to show that closed string amplitudes with
$b$ boundaries are identical to closed string amplitudes with $b$ additional 
insertions of a particular physical closed string state. 
We perform an explicit computation for the annulus, and argue on the basis of open and 
closed string field theory for higher order amplitudes.
We also discuss possible subtleties in the prescription related to
collisions of boundaries and insertions, and argue that they are harmless.
This verifies the proposal that a periodic
array of D-branes in imaginary time corresponds to a pure closed string background.}

\keywords{D-branes, S-branes, open/closed string duality}

\begin{document}

\section{Introduction}

Open/closed string duality has been a powerful tool in understanding string theory
and gauge theory.
The most basic form of this relation originates from the dual descriptions
of the vacuum cylinder amplitude, as a one-loop open string vacuum energy on one
hand, and as a tree-level closed string exchange amplitude on the other.
The AdS/CFT correspondence is another kind of open/closed string duality,
whereby an open string theory on a collection of D-branes
is dual to a closed string theory in some non-trivial geometry with fluxes.
Recently, another type of open/closed string duality has been revealed.
Gaiotto, Itzhaki and Rastelli have shown that closed string disk scattering
amplitudes from a periodic array of D-branes in imaginary time are identical
to purely closed string sphere amplitudes with an additional insertion
of a particular physical closed string state \cite{a:GIR}.
This led to the proposal that
an array of D-branes in imaginary time corresponds to a
particular pure closed string background, and that perhaps any configuration
of D-branes which are localized in imaginary (or complex) time corresponds
to some pure closed string background.

Arrays of D-branes in imaginary time arise in the study of unstable D-brane
creation and evaporation \cite{a:GS1,a:SEN,a:Strominger,a:MSY,a:LLM}, 
a process also known as a spacelike brane, or S-brane.
An S-brane is described by adding to the flat space CFT the marginal
boundary deformation
\be
\lambda\int dt \cosh (x_0(t)) \;.
\ee
This is related to the boundary sine-Gordon theory by a Wick rotation
$x_0=ix$. In that case the deformation $\lambda\int dt\,\cos(x(t))$
interpolates between a Neumann boundary condition at $\lambda=0$
and a Dirichlet boundary condition at $\lambda=1/2$ \cite{Callan_Maldacena,Polchinski}.
The latter therefore corresponds to a periodic array of D-branes in $x$.
Upon Wick rotation we therefore formally get an array of D-branes in
imaginary time when $\lambda=1/2$. This is a time-dependent background
which describes the creation at $x_0=0$,
and immediate decay, of an unstable D-brane. It is therefore natural
to expect that it can be described purely in terms of closed strings.

The authors of \cite{a:GIR} derived a prescription for computing
disk amplitudes for D-branes in imaginary time using a specific
prescription for analytically  continuing amplitudes for D-branes
in real space.
Applying this to the two-point (and more generally $n$-point)
disk amplitude shows that the only
contribution comes from the limit where the size of the boundary shrinks
to a point, and that this corresponds to an additional insertion of
a specific physical closed string state $|W\rangle$. Assuming this holds for amplitudes
with any number of boundaries, namely that the only contribution comes from the limit
where all the boundaries shrink, and in that limit each boundary is replaced
with an insertion of $|W\rangle$, the insertion will exponentiate, implying
that this is really a closed string {\em background}.

The aim of the present paper is to extend this result to higher order
amplitudes, including an arbitrary number of boundaries and genus.
We will show that any amplitude which involves a periodic array of D-branes in
imaginary time is identical to a purely closed string amplitude where
the boundaries are replaced by additional insertions of the same closed string
state $|W\rangle$. This will complete the verification of the proposal that the D-brane
array corresponds to a pure closed string background.

Our paper is organized as follows. In section 2 we first review
the prescription for calculating disk amplitudes for an array of D-branes
in imaginary time presented in \cite{a:GIR}, and then
generalize for amplitudes with any number of boundaries
(in the Appendix we show
how the prescription can be extended to an arbitrary configuration of imaginary branes).
In section 3 we apply the general prescription to the
two-point annulus amplitude, and show that it is identical
to the four-point sphere amplitude with two $|W\rangle$s.
In section 4 we will address the question of
additional contributions to the amplitude from collision singularities.
We will argue that the prescription can be chosen in such a way
that these additional singularities do not contribute.
General amplitudes will de discussed in section 5, where, relying
on formulations of closed and open/closed string field theory, we will show
that an amplitude with $b$ boundaries and $n$ closed string insertions
is identical to an amplitude with no boundaries and $n+b$ insertions.

\section{General Prescription}\label{GenCons}

\subsection{Review of the disk case}

We begin with a brief review of the prescription for the disk amplitudes \cite{a:GIR}.
If $\widetilde A$ is a disk amplitude of closed strings with a single D-brane
located at a spatial position $x=0$, then the amplitude with a periodic array
at $x = (n+\half)a$ is given by
\be
\label{disc}
\widetilde S(P)=\sum^{\infty}_{n=-\infty}
\widetilde A(P)e^{i(n+\half)aP}=\widetilde A(P)
\sum^{\infty}_{n=-\infty}(-1)^n2\pi \delta(Pa-2\pi n)\;,
\ee
where we have suppressed all the kinematic variables except the total momentum
in the $x$ direction $P$.
The amplitude for an array of D-branes in imaginary time is then defined
by the Wick rotation $x\to -ix^0$ and $P\to iE$, with $x^0$ and $E$ real.
This gives
\be
S(E)=\widetilde A(iE)\sum^{\infty}_{n=-\infty}(-1)^n2\pi \delta(iEa-2\pi n).
\ee
Naively this vanishes, however $\widetilde A(iE)$ may blow up for some values of $E$
and yield a non-zero result.
One has to provide a prescription for computing this quantity, which is
essentially a prescription for performing the analytic continuation
from real space to imaginary time. The prescription given in \cite{a:GIR} is as
follows.
Consider the Fourier transform of the original amplitude
\be
 \widetilde{G}(x) = \int_{-\infty}^{\infty} dP \, e^{iPx} \widetilde A(P)
\sum^{\infty}_{n=-\infty}(-1)^n2\pi \delta(Pa-2\pi n)\;.
\ee
Using the residue theorem this can be written as
\be
 \widetilde{G}(x) = \frac{1}{2i} \oint_{\cal C} dP\, e^{iPx}
\frac{\widetilde{A}(P)}{\sin(aP/2)}\;,
\ee
where the contour $\cal{C}$ is shown in figure 1. An assumption is then made
that $\widetilde{A}(P)$ is an analytic function, with poles or cuts only
along the imaginary $P$ axis.
The contour $\cal C$ can therefore be deformed to
a contour ${\cal C}'$, which implies that
\be
 S(E) = \frac{\mbox{Disc}_E[\widetilde{A}(iE)]}{2\sinh(aE/2)}\;,
\label{GIR_pres}
\ee
where
\be
 \mbox{Disc}_E[f(E)] \equiv -i[f(E+i\epsilon) - f(E-i\epsilon)] \;.
\ee

The above prescription was then applied to the two-point disk amplitude.
We repeat the computation in the boundary state approach, since that is the one
which is easiest to generalize to higher order. This amplitude has a single
modulus, which in the parameterization of \cite{a:GIR} corresponds to the
radius of the boundary $\rho$. The disk corresponds to $|z|\geq\rho$, and
the two insertions are fixed at $z=1$ and $z=\infty$.
The amplitude is therefore given by
\be
 \widetilde{A}(p_1,p_2) = \int_0^1 \frac{d\rho}{\rho}\,
 \langle 0 | V(p_1;\infty) V(p_2;1) (b_0+\tilde{b}_0)
\rho^{L_0+\tilde{L}_0} |\widetilde{B}\rangle \;,
\ee
where $|\widetilde{B}\rangle$ is a boundary state at radius 1 corresponding to a
D-brane located at $x=0$, and the operator $\rho^{L_0+\tilde{L}_0}$ propagates the
boundary state from radius 1 to radius $\rho\leq 1$.
Inserting a complete set of closed string states and integrating over $\rho$
gives
\be
 \widetilde{A}(p_1,p_2) = \int dk\,\sum_i
 \langle 0 | V(p_1;\infty) V(p_2;1)|k,i\rangle
 \frac{1}{\frac{k^2}{ 2} + 2l_i} \langle k,i|(b_0+\tilde{b}_0)|\widetilde{B}\rangle\;,
\ee
where $l_i$ is the level of the state $|k,i\rangle$.
Applying the prescription (\ref{GIR_pres}) then gives the three-point sphere amplitude
\be
 S(p_1,p_2) = \langle 0 | V(p_1;\infty) V(p_2;1)|W_E\rangle\;,
\ee
where $|W_E\rangle$ is a
physical (ghost number two) closed string state defined in terms
of the Wick-rotated boundary state\footnote{The boundary state itself
is not a physical state since it has ghost number 3.}
\begin{eqnarray}
|{W_E}\rangle & \equiv & \int dk\sum_i|k,i\rangle\frac{\delta(k^2/2+2l_i)}
{2\sinh(Ea/2)}\langle k,i|(b_0+ \tilde b_0)|B\rangle \nonumber\\
 & = & \frac{\delta(L_0+\tilde L_0)}{2\sinh(Ea/2)}(b_0+\tilde b_0)|B\rangle \;.
\end{eqnarray}
\FIGURE[ht]{
\let\picnaturalsize=N
\def\picsize{3in}
\def\picfilename{Contours.eps}
\ifx\nopictures Y\else{\ifx\epsfloaded Y\else\input epsf \fi
\let\epsfloaded=Y
\centerline{\ifx\picnaturalsize N\epsfxsize \picsize\fi
\epsfbox{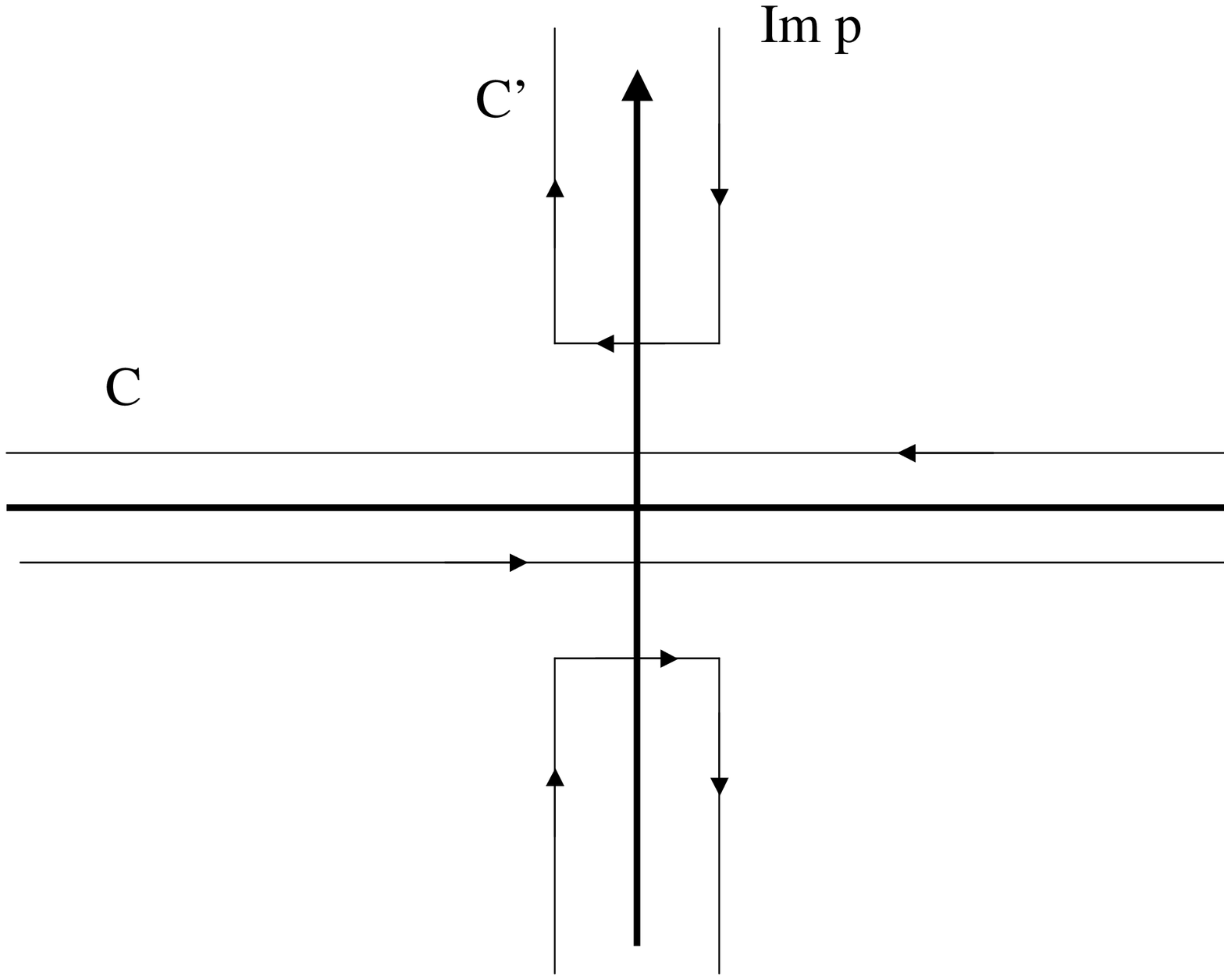}}}\fi
\caption{Integration contours}
\label{contours}
}

\subsection{Any number of boundaries}

Consider now an amplitude with $b$ boundaries.
Here we need to take into account the possibility of associating
different boundaries to different D-branes in the spatial array.
Denote by $\widetilde A_{n_1...n_b}$ the amplitude with the
first boundary on the $n_1$th brane, the second on the $n_2$th brane, etc.
The total amplitude is then given by
\be
\widetilde S=\frac{1}{b!}\sum_{\{n_j\}}\widetilde A_{n_1...n_b}\;.
\ee
In terms of the boundary-momentum-space amplitude $\widetilde{A}(k_1,\ldots,k_b)$
this becomes
\begin{eqnarray}
\widetilde S & = & \frac{1}{b!} \sum_{\{n_j\}}
\int \prod_{j=1}^{b} \left[dk_j\, e^{iak_j(n_j + 1/2)}\right]
\widetilde{A}(k_1,\ldots,k_b)\, \delta\left(\sum_{j=1}^b k_j-P\right) \nonumber\\
 &= & \frac{1}{b!} \int \prod_{j=1}^b dk_j\,
\widetilde{S}(k_1,\ldots,k_b)\, \delta\left(\sum_{j=1}^bk_j-P\right)\;,
\end{eqnarray}
where $P$ is the total momentum in the direction of the array, and
\be
 \widetilde{S}(k_1,\ldots,k_b) = \widetilde{A}(k_1,..,k_b)
 \prod_{j=1}^{b} \sum_{n_j=-\infty}^{\infty} (-1)^{n_j}\, 2\pi \delta(k_j a-2\pi n_j)\;.
\ee
For $b=1$ this reduces to the disk amplitude (\ref{disc}).
Consider the Fourier transform of this expression with respect to 
the boundary momentum $k_1$,
\begin{eqnarray}
\widetilde G(x,k_2,\ldots,k_b) &=& \int dk_1\, e^{ik_1x} \widetilde{S}(k_1,\ldots,k_b)
 \nonumber\\
 &=& \frac{1}{2i} \oint_{\cal C} dk_1\, e^{ik_1x}\,
\frac{\widetilde{A}(k_1,\ldots,k_b)}{ \sin(ak_1/2)}\prod_{j=2}^{b}
\sum_{n_j=-\infty}^{\infty}2\pi \delta(k_ja-2\pi n_j).
\ee
As before, we would like to deform the contour to ${\cal C}'$.
Unlike the single boundary case however, we cannot assume that
$\widetilde{A}(k_1,\ldots,k_b)$ has singularities only along the
imaginary $k_1$ axis. Generically it will have singularities also
along a finite number of lines parallel to the imaginary axis.
Schematically, these arise from terms of the form
$$
\frac{1}{(k_1-q)^2+a^2}\;,
$$
where $q$ is some combination of the other boundary momenta.
The deformed contour will then contain a number of contours like ${\cal C}'$.
Upon Fourier transforming back we obtain an expression for
$\widetilde S(k_1,\ldots,k_b)$ in terms of the contour integrals.
However, since we are interested in the Wick-rotated amplitude with
$k_1=iE_1$ and $E_1$ real,
only the contour on the imaginary axis contributes,
and the result is
\be
\widetilde{S}(iE_1,k_2,\ldots,k_b)=
\frac{\mbox{Disc}_{E_1}\widetilde{A}(iE_1,k_2,\ldots,k_M)}{
2\sinh(aE_1/2)}
\prod_{j=2}^{b}\sum_{n_j=-\infty}^{\infty}2\pi \delta(k_ja-2\pi n_j) \;.
\ee
Repeating this procedure for the rest of the boundary momenta $k_2,\ldots,k_b$
we obtain
\be
S(E_1,\ldots,E_b)=
\frac{\mbox{Disc}_{E_1}[\mbox{Disc}_{E_2}
[...\mbox{Disc}_{E_b}[\widetilde{A}(iE_1,\ldots,iE_b)]...]]}{
\prod_{j=1}^b 2\sinh(aE_j/2)} \;.
\label{prescription}
\ee
The total amplitude is then given by
\be
 S(E) = \frac{1}{b!}\int\left[\prod_{i=1}^{b}dE_i\right]
  \delta\left(\sum_{k=1}^b E_i-E\right)
   S(E_1,\ldots,E_b)\;,
\label{total_amplitude}
\ee
where $E=-iP$ is the total energy.
This will be our prescription for calculating amplitudes off a periodic
array of D-branes in imaginary time.
For more general brane configurations see Appendix ~\ref{app1}.
The prescription of ~\cite{a:GIR} is recovered by setting $b=1$.

\section{Annulus two-point amplitude}

Let us now apply our prescription to the two-point annulus amplitude.
We first need to choose a convenient parameterization of the annulus.
Consider the usual parameterization of the annulus (Fig.~\ref{an_moduli}a):
the external radius is $1$, the internal radius is $\rho$, one insertion
is located on the real axis (using rotational symmetry) at $r$, and
the other is located at $z$. The modulus $\rho$  satisfies
$0\leq \rho\leq r\leq 1$.
We would like to integrate over the radii and fix as many insertions as possible.
Consider then the  transformation $z\to z/r$. This fixes the position of the first
insertion at $1$, and gives
\be
\rho_{out}=\frac{1}{r}\;,\;
\rho_{in}=\frac{\rho}{r}\;,\;
z'=\frac{z}{r}
\ee
for the external radius, internal radius, and position of the second insertion,
respectively (Fig.~\ref{an_moduli}b). These are the (four) real moduli of the annulus with 
two insertions.
The radii satisfy $1\leq \rho_{out}\leq \infty$ and $0\leq \rho_{in}\leq 1$.
\FIGURE[ht]{
\let\picnaturalsize=N
\def\picsize{3.5in}
\def\picfilename{Annulus.eps}
\ifx\nopictures Y\else{\ifx\epsfloaded Y\else\input epsf \fi
\let\epsfloaded=Y
\centerline{\ifx\picnaturalsize N\epsfxsize \picsize\fi
\epsfbox{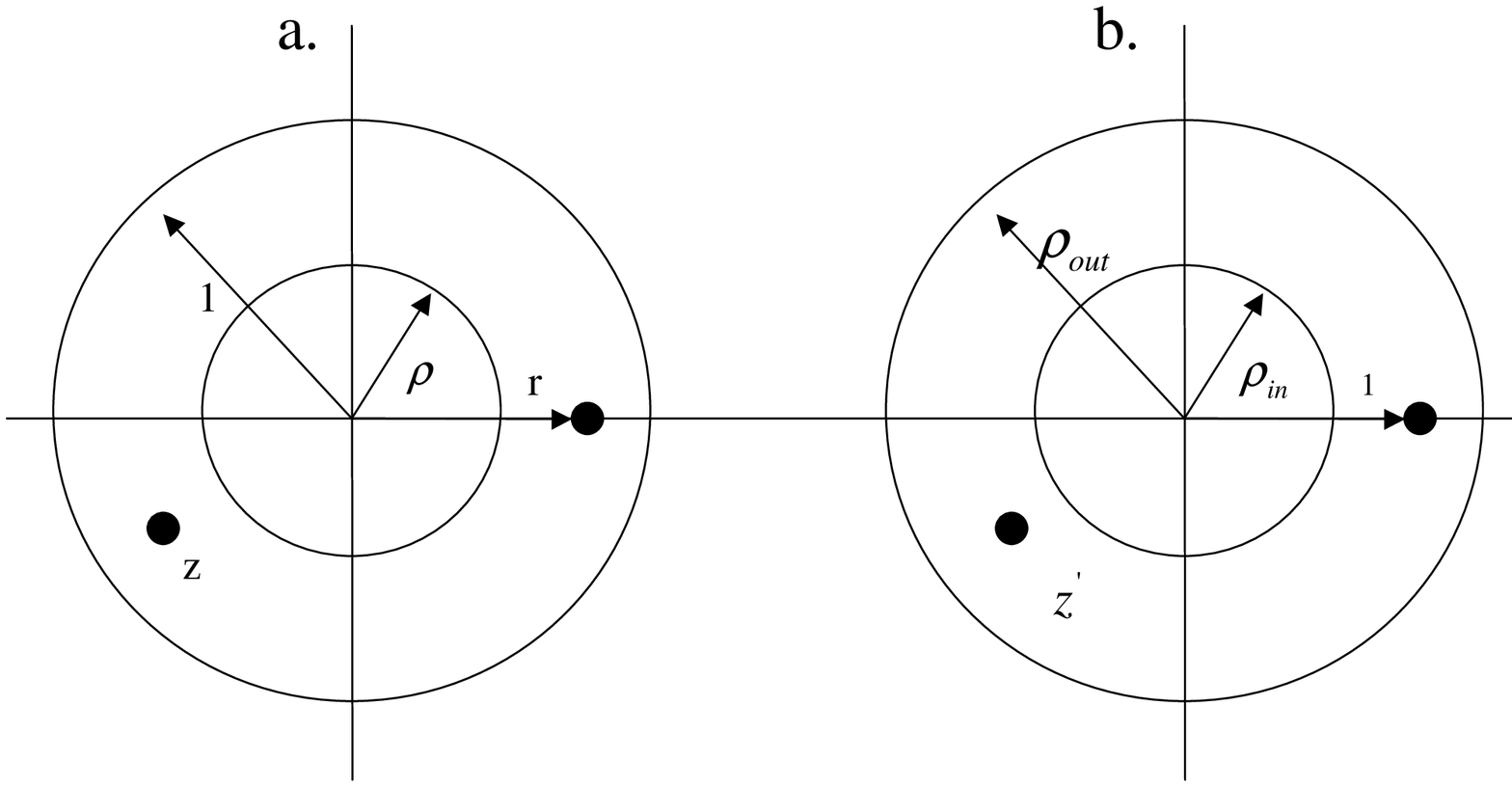}}}\fi
\caption{(a) The usual parametrization of the annulus, and
(b) the new parametrization of the annulus.}
\label{an_moduli}
}
The amplitude is then given by
\begin{eqnarray}
\widetilde{A}(p_1,p_2)&=&\int d^2z'\int_0^1\frac{d\rho_{in}}{\rho_{in}}
\int_1^\infty\frac{d\rho_{out}}{\rho_{out}}\\
&&\langle \widetilde{B}|
\rho_{out}^{-L_0-\tilde L_0}(b_0+\tilde{b}_0){V}_1(p_1;1,1){V}_2(p_2;z',\bar z')
(b_0+\tilde b_0)\rho_{in}^{L_0+\tilde L_0}|\widetilde{B}\rangle\;,\nonumber
\end{eqnarray}
Inserting complete sets of states, and integrating over the radii gives
\be
 \widetilde{A}(p_1,p_2) = \int dk_1 dk_2 \, \widetilde{A}(k_1,k_2;p_1,p_2) \;,
\ee
where
\begin{eqnarray}
\label{Amplit}
\widetilde{A}(k_1,k_2;p_1,p_2) &=& \sum_{i,j} \int d^2z'\,
\langle \widetilde{B}|(b_0+\tilde b_0)|k_1,i\rangle\langle k_1,i|
\frac{1}{\frac{k_1^2}{2}+2l_i}
\\
&&  {V}_1(p_1;1,1){V}_2(p_2;z',\bar z')\frac{1}
{\frac{k_2^2}{2}+2l_j}|k_2,j\rangle\langle k_2,j|(b_0+\tilde b_0)
 |\widetilde{B}\rangle \;.\nonumber
\end{eqnarray}
This is the amplitude corresponding to boundary momenta $k_1$ and $k_2$,
to which we will apply our general prescription (\ref{prescription}).
It is apparent that this amplitude has poles on the imaginary $k_1$ and
$k_2$ axes, which will contribute to the discontinuities.
Additional singularities can arise from the integral over $z'$,
but as we shall argue in the next section, their contribution can
be neglected in (\ref{prescription}). We therefore obtain for the total amplitude
\be
S(p_1,p_2)=\frac{1}{2}\int dE_1\int d^2z'\, \langle {W_{E_1}}|
{V}_1(p_1;1,1){V}_2(p_2;z',\bar z')|{W_{E-E_1}}\rangle\;.
\ee
As anticipated, we obtain the amplitude for four closed string states on
the sphere. We can also express this as
\be
 S(p_1,p_2) = \frac{1}{2}\langle W(\infty,\infty) {V}_1(p_1;1,1) {V}_2(p_2;z',\bar z')
              W(0,0) \rangle \;,
\ee
where $W(z,\bar{z})$ is the vertex operator corresponding to the
state $\int dE\, |W_E\rangle$.

\section{Collisions of boundaries and insertions}

Before we generalize this result to higher orders,
let us turn our attention to the possible contributions
of other singularities.
Amplitudes with boundaries and bulk insertions can have four types
of singularities, arising respectively from a shrinking boundary,
the collision of two vertex operators,
the collision of a vertex operator and a shrinking boundary and
the collision of two shrinking boundaries.
The singularities are all poles, and are associated with the emission of
an on-shell closed string.
In the previous section we assumed that only the shrinking-boundary
singularities contribute to the discontinuity in the annulus amplitude.
In the next section we will assume that the same holds for
the discontinuity of a general amplitude.
This is a crucial condition for producing the correct closed
string amplitudes.
Here we will show that it is actually implied by the choice
of analytic continuation.

Consider first a shrinking-boundary singularity.
Apart from analytic factors, the amplitude with a shrinking
boundary is basically of the form
\be
\widetilde{A}_B(k) = \frac{1}{k^2 + m^2} \;,
\label{B}
\ee
where $k$ is the momentum of the shrinking boundary.
This amplitude was used as an illustration of a disk amplitude in
\cite{a:GIR}. Their strategy for computing the amplitude for an array of D-branes in
imaginary time was to Fourier-transform to position space,
sum over an array of D-branes in space, analytically continue to
imaginary time and finally transform back to momentum (energy) space.
In position space we get
\be
 \widetilde{G}_B(x) = \int dk\, e^{ikx} \widetilde{A}_B(k)
  = \frac{\pi}{m} e^{-m|x|}\;.
\ee
Summing over the array then gives
\be
\widetilde{G}_{B,array}(x) = \frac{\pi}{ m}
\sum_{n=-\infty}^{\infty}e^{-m|x+a(n+\half)|}
 =\frac{\pi}{ m}\, \frac{{\textit f}(mx)}{\sinh(ma/2)}\;,
\ee
where the function $f(mx)$ is defined as $\cosh(mx)$ for $-\half a\leq x\leq \half a$,
and is periodic with period $a$. This expression is not analytic, however it is analytic
within each period. A natural prescription for analytically continuing it to
the imaginary axis is to focus on the branch around the origin,
{\em i.e.} $f(mx)=\cosh(mx)$ for all $x$, in which case
\be
 G_{B,array}(x^0) = \frac{\pi}{ m}\,\frac{\cos(mx^0)}{\sinh(ma/2)}\;,
\ee
and in momentum space
\be
 S_B(E) = \frac{\pi}{ 2m\sinh(ma/2)}\left[\delta(E-m)+\delta(E+m)\right]
  = \frac{\pi\delta(E^2-m^2)}{\sinh(ma/2)}\;.
\ee
This agrees precisely with the general prescription (\ref{prescription}) applied 
to (\ref{B}).

The other singularities, corresponding to the collision
of insertions and shrinking boundaries, arise
when the sum of the momenta of the two
colliding elements is on-shell for some closed string state.
Thus amplitudes with two colliding insertions, a colliding insertion
and shrinking boundary and two colliding shrinking boundaries are
respectively of the qualitative form
\begin{eqnarray}
 \widetilde{A}_{VV}(p_1,p_2) &=& {1\over (p_1+p_2)^2 + m^2}
 \label{VV}\\
 \widetilde{A}_{VB}(k,p) &=& {1\over k^2 + m^2}\,{1\over (k+p)^2+m^{\prime 2}}
 \label{VB}\\
  \widetilde{A}_{BB}(k_1,k_2) &=& {1\over k_1^2 + m_1^2}\,{1\over k_2^2 + m_1^2} \,
 {1\over (k_1+k_2)^2 + m^2}
 \label{BB}\;.
\end{eqnarray}
In the case of the annulus two-point amplitude, for example, there is a singularity of
the first type when $z'\to 1$ and $(p_1+p_2)^2$ is on-shell, and a
singularity of the second type when $\rho_{in}\to 0$, $z'\to 0$
and $(p_1+k_1)^2$ is on-shell (and another when $\rho_{out}\to \infty$,
$z'\to \infty$ and $(p_1+k_2)^2$ is on-shell).
Boundary-boundary singularities will arise when there are more boundaries.
As explained in \cite{a:GIR}, the poles coming from collisions of vertex
operator insertions (\ref{VV}) are harmless,
since one can always define the amplitude by analytically continuing
the external momenta $p_i$ off-shell away from the singularity.
The same argument does not work for the boundary momenta $k_j$,
since the amplitude contains an integral over all their possible values.
We therefore need a different argument for the insertion-boundary
and boundary-boundary poles.\footnote{For the special case of the disk,
the insertion-boundary pole is harmless for the same reason as above.
Momentum conservation implies that if $p_1+k$ is on-shell,
then so is the sum of all the other external momenta $\sum_{i=2}^n p_i$, 
and the latter can be analytically
continued off-shell in the amplitude.}

Let us compute directly the amplitudes for D-branes in imaginary time
derived from (\ref{VB}) and (\ref{BB}) using the same strategy as above.
In both cases there are poles from shrinking boundaries and poles from
collisions. We will show that only the former contribute in our prescription.
In position space we get
\begin{eqnarray}
 \widetilde{G}_{VB}(x) &=& {\pi\over 2mm'}
 \int_{-\infty}^{\infty} dy\, e^{-m'|y|-ipy}e^{-m|x-y|}\\
 \widetilde{G}_{BB}(x_1,x_2) &=& {\pi^2\over 2m_1m_2m}
  \int_{-\infty}^{\infty}dy\, e^{-m_1|y-x_1|}e^{-m_2|y-x_2|}e^{-m|y|}.
\end{eqnarray}
Summing over the array then gives
\begin{eqnarray}
 \widetilde{G}_{VB,array}(x) &=&  {\pi\over 2mm'}
  \int_{-\infty}^{\infty} dy\, e^{-m'|y|-ipy}\,
\frac{{\textit f}(m(x-y))}{\sinh(ma/2)}
 \label{VBarray}\\[5pt]
\widetilde{G}_{BB,array}(x_1,x_2) &=& {\pi^2\over 2m_1m_2m} \int_{-\infty}^{\infty}dy\,
 {f(m_1(y-x_1)) \over \sinh(m_1a/2)}\,
 {f(m_2(y-x_2)) \over \sinh(m_2a/2)}\, e^{-m|y|},
\label{BBarray} 
\end{eqnarray}
where $f$ is the same periodic function as before.
We need to specify again how to analytically continue $f$ to the imaginary axis.
The most natural choice is to use the same prescription as before, namely
to replace $f(m(x-y))$ with $\cosh(m(x-y))$ in both (\ref{VBarray})
and (\ref{BBarray}).
Wick rotating and then transforming back to momentum space we finally get
\begin{eqnarray}
 S_{VB}(E) &=& {\pi \delta(E^2-m^2) \over\sinh(ma/2)}\,
             {1 \over (E -ip)^2-m'^2}\\[5pt]
 S_{BB}(E_1,E_2) &=& \frac{\pi\delta(E_1^2-m_1^2)}{\sinh(m_1a/2)}\,
 \frac{\pi\delta(E_2^2-m_2^2)}{\sinh(m_2a/2)}\,
 \frac{1}{(E_1+E_2)^2-m^2}\;,
\end{eqnarray}
in complete agreement with our assertion that the general prescription
(\ref{prescription}) receives contributions only from
shrinking boundaries.

The generalization to an arbitrary number of colliding vertices and shrinking
boundaries is straightforward. The singular part of the amplitude is
\be
 \prod_{j=1}^b\frac{1}{k_j^2+m_j^2} \,
 \frac{1}{(\sum_{i=j}^b k_j + \sum_{i=1}^n p_i)^2 + m^2}\;,
\label{general_vb}
\ee
which gives
\be
{\pi^b\over 2m\prod_{j=1}^bm_j}\int_{-\infty}^{\infty}dy\, \prod_{j=1}^b\,
\frac{{\textit f}(m_j(y-x_j))}{\sinh(m_ja/2)}e^{-m|y|-iy\sum_i p_i},
\ee
for the position space amplitude of the array.
Applying the same prescription gives
\be
  \prod_{j=1}^b\, {\pi\delta(E_j^2-m_j^2)\over\sinh(m_ja/2)}\,
  {1\over (\sum_{j=1}^b E_j)^2 - m^2}\;,
\label{general_vb_total}
\ee
which is precisely what one would get by computing the discontinuities
of (\ref{general_vb}) from the shrinking boundaries.

In summary, only the poles coming from boundaries shrinking to points contribute
in our prescription (\ref{prescription}). The other poles remain in the
resulting amplitudes, {\em e.g.} (\ref{general_vb_total}), as they should
for closed string amplitudes.

\section{General amplitudes}\label{sectVac}

A sphere amplitude with $b$ boundaries and $n$ closed string vertices has $3b$
boundary moduli and $2n$ closed string moduli.
The CKVs of the sphere can be used to fix the positions of three boundaries,
leaving integrations over
the remaining $b-3$ complex boundary positions, $b$ boundary radii and
$n$ complex positions of the vertices. The dimensionality of the moduli space
is the same as for an $(n+b)$-punctured sphere with
additional integrations over the boundary sizes.
More generally, a genus $g$ amplitude with $n$ closed string vertices
and $b$ boundaries has $(6g-6+2n+2b)$ real moduli, which is the same same
as a genus $g$ amplitude with $n+b$ vertices and no boundaries plus
the extra boundary size integrals.

We have shown that the general prescription for computing any amplitude
involving an array of D-branes in imaginary time (\ref{prescription})
receives a contribution only in the limit where all the boundaries
have a vanishing size. To see that this corresponds precisely to a closed
string amplitude we still need to show that the moduli space of
the resulting punctured Riemann surface is covered completely and without
overlap. In other words we require a slicing of the moduli space
of punctured Riemann surfaces with boundaries, and a slicing of the moduli
space of punctured Riemann surfaces without boundaries, such that the former
reduces to the latter when all the boundaries shrink.
Such slicings have been constructed by Zwiebach in
the context of closed string field theory \cite{a:zwie1}, and open/closed string
field theory \cite{a:zwie2}. We will use these to demonstrate
the above reduction. Let us first review (very) briefly Zwiebach's results.

\subsection{Zwiebach's closed and open/closed SFTs}

String field theory can be viewed as a set of vertices, which can
be used to build any string amplitude using Feynman rules.
One of the main difficulties in doing this for the closed string (covariantly)
is the need to cover the moduli space of every diagram completely
and exactly once. This problem was solved by Zwiebach for the purely closed bosonic
string in \cite{a:zwie1}. The closed string vertices $\langle V_{g,n}|$
are defined in terms of regions of the moduli space ${\cal V}_{g,n}$
satisfying two properties. The first is that
every diagram formed with the vertices has a minimal area metric, subject to
the constraint that every non-contractible closed curve on it has length
greater than or equal to $2\pi$.
This implies that every point in the moduli space is covered at most once.
It also implies that the vertices have stubs of length $\pi$.
In addition, the vertices satisfy recursion relations,
which can be expressed concisely as
\be
\partial{\cal V}+\half\left\{{\cal V},{\cal V}\right\}+\hbar\Delta{\cal V}=0\;,
\ee
where ${\cal V}$ represents the formal sum of the vertices ${\cal V}_{g,n}$.
The operation $\{\,,\}$ corresponds to sewing two punctures located on different
vertices, $\Delta$ corresponds to sewing two punctures on the the
same vertex, and $\partial$ is the boundary operator.
The specific relation for the moduli space of a given topology ${\cal M}_{g,n}$
is recovered by considering all terms of the same topology.
This relation guarantees that every point in the moduli space is covered.
For example, for the four-punctured sphere one gets
\be
 \partial {\cal V}_{0,4} + {1\over 2}\{{\cal V}_{0,3},{\cal V}_{0,3}\} = 0 \;.
\ee
This shows that the four-punctured sphere is covered fully and without overlap
by four regions,
corresponding to the three diagrams formed by two 3-string vertices and
a propagator, and the 4-string vertex.

In \cite{a:zwie2} it was shown how to extend this framework to open and closed
string field theory. One defines string vertices which include $b$ boundaries
and $m$ boundary insertions, ${\cal V}_{g,n}^{b,m}$, using the same properties
as above, where the definitions of the anti-bracket and $\Delta$ are extended
to incorporate the sewing of open string vertices as well.
The purely closed string vertices are then the same as before,
namely ${\cal V}_{g,n}^{0,0}={\cal V}_{g,n}$.

\subsection{General amplitudes of imaginary-time D-branes}

Consider a genus $g$ amplitude with $n$ closed string insertions and $b$
boundaries (and no open string insertions).
What we are trying to show is that if the boundaries correspond to an array
of D-branes in imaginary time, this amplitude is identical to a
genus $g$ amplitude with $n+b$ insertions and no boundaries.
Qualitatively they are the same,
since the general prescription picks out the
limit of vanishing boundaries, and our arguments in section 4 show
that other limits do not contribute.
However one still needs to show that the closed string moduli space
is properly covered in this limit.

The moduli space of the amplitude with boundaries is divided into
several regions, each corresponding to a different diagram built from
vertices with boundaries and propagators.
In particular, there exists a region of the moduli space in which
the diagram is built by connecting $b$ string-boundary vertices
${\cal V}_{0,1}^{1,0}$ to the genus $g$ $(n+b)$-string vertex
${\cal V}_{g,n+b}^{0,0}$ using $b$ propagators.\footnote{The string-boundary vertex
corresponds to the boundary state $|B\rangle$.}
The region corresponds to varying the length of each propagator from
zero to infinity. When a particular propagator has zero length we get
the stitch $\{{\cal V}_{0,1}^{1,0},{\cal V}_{g,n+b}^{0,0}\}$, which matches on
to the boundary of the region built with ${\cal V}_{g,n+b-1}^{1,0}$.
The limit in which all the propagators have infinite length corresponds
to the diagram where all the boundaries have zero size.
As we argued in the previous section, this is the only relevant part
of the amplitude for the general prescription.
The above region is therefore the only relevant region, and since
${\cal V}_{g,n+b}^{0,0}={\cal V}_{g,n+b}$ we conclude that the slicing
of the closed string moduli space is correctly recovered.

\FIGURE[ht]{
\let\picnaturalsize=N
\def\picsize{2.5in}
\def\picfilename{General.eps}
\ifx\nopictures Y\else{\ifx\epsfloaded Y\else\input epsf \fi
\let\epsfloaded=Y
\centerline{\ifx\picnaturalsize N\epsfxsize \picsize\fi
\epsfbox{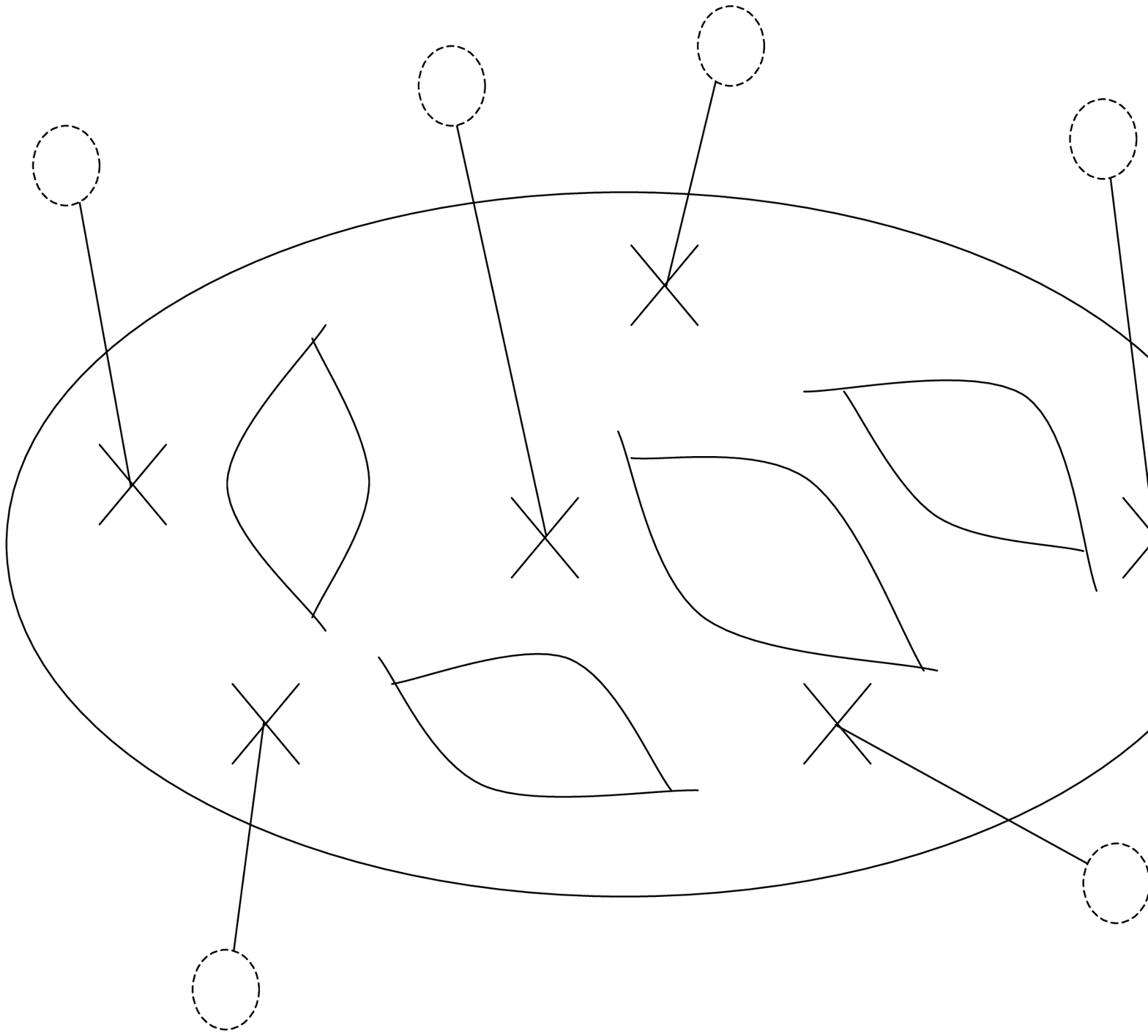}}}\fi
\caption{A general amplitude with boundaries introduced via closed string boundary vertices.}
\label{_3_amp_bound}
}

The only thing which remains to be shown is that the $b$ additional insertions
correspond to the state $|W\rangle$.
To this end we will compute the contribution of the above region
to the amplitude explicitly, and then apply the prescription (\ref{prescription}).
Let us denote the positions of the closed string insertions by $z_i$,
the positions of the boundaries by $\zeta_j$, and their radii by $\rho_i$.
The lengths of the propagators correspond to the radii of the boundaries:
infinite length corresponds to zero radius, and zero length
to some maximal radius $\rho_i^{max}$.
The amplitude is therefore given by (we omit the ghosts for the sake of clarity)
\be
\widetilde{A}_{g,n}^{b,0}(p_1\ldots,p_n) =
 \langle V_{g,n+b}^{0,0}|
  \prod_{i=1}^n |V_i(p_i;z_i,\bar{z}_i)\rangle
  \prod_{j=1}^b \int_0^{\rho_j^{max}} {d\rho_j\over\rho_j}\,
   \rho_j^{L_0^{(j)}+\tilde{L}_0^{(j)}}\,|B_j(\zeta_j,\bar{\zeta}_j)\rangle .
\ee
The integrals over the position and genus moduli are
included in the closed string vertex $\langle V_{g,n+b}^{0,0}|$.
The range of integration of each radius will depend in general on
all the other moduli. However, as we shall see, our result is
independent of the details of this dependence.
Performing the $\rho_b$ integral gives
\begin{eqnarray}
 \lefteqn{\widetilde{A}_{g,n}^{b,0}(p_1\ldots,p_n) =} \nonumber\\
 & & \langle V_{g,n+b}^{0,0}|
  \prod_{i=1}^n |V_i(p_i;z_i,\bar{z}_i)\rangle
   \prod_{j=1}^{b-1} \int_0^{\rho_j^{max}} {d\rho_j\over\rho_j}\,
   \rho_j^{L_0^{(j)}+\tilde{L}_0^{(j)}}\,|B_j(\zeta_j,\bar{\zeta}_j)\rangle
   {(\rho_b^{max})^{L_0^{(b)}+\tilde{L}_0^{(b)}}\over L_0^{(b)} + \tilde{L}_0^{(b)}}
   |B_b(\zeta_b,\bar{\zeta}_b)\rangle . \nonumber\\
\end{eqnarray}
Inserting a complete set of momentum states for the $b$th boundary, and
computing the discontinuity with respect to $E_b$ then gives
\be
 -2\pi \langle V_{g,n+b}^{0,0}|
  \prod_{i=1}^n |V_i(p_i;z_i,\bar{z}_i)\rangle
   \prod_{j=1}^{b-1} \int_0^{\rho_j^{max}} {d\rho_j\over\rho_j}\,
   \rho_j^{L_0^{(j)}+\tilde{L}_0^{(j)}}\,|B_j(\zeta_j,\bar{\zeta}_j)\rangle
   \delta(L_0^{(b)} + \tilde{L}_0^{(b)})
   |B_b(\zeta_b,\bar{\zeta}_b)\rangle . \nonumber\\
\ee
That the result is independent of $\rho_b^{max}$ follows from
$\mbox{Disc}(\rho^E/E) = -2\pi\rho^E\delta(E)=-2\pi\delta(E)$.
Repeating this procedure for the remaining radii we obtain
\be
 (-2\pi)^b\langle V_{g,n+b}^{0,0}|
  \prod_{i=1}^n |V_i(p_i;z_i,\bar{z}_i)\rangle
   \prod_{j=1}^{b} \delta(L_0^{(j)}+\tilde{L}_0^{(j)})
   |B_j(\zeta_j,\bar{\zeta}_j)\rangle\;.
\ee
The total amplitude which follows from (\ref{prescription}) and (\ref{total_amplitude})
is therefore given by
\be
 S(p_1,\ldots,p_n) = {1\over b!}
 \langle V_{g,n+b}^{0,0}| \prod_{i=1}^n |V_i(p_i;z_i,\bar{z}_i)\rangle
   \prod_{j=1}^{b-1} \int dE_j\,
   |W_{E_j}\rangle\, |W_{E-\sum_{j=1}^{b-1} E_j}\rangle \;,
\ee
where $E$ is the total energy of the closed string insertions.
This can also be expressed as
\be
 S(p_1,\ldots,p_n) = {1\over b!} \int_{{\cal V}_{g,n+b}}
 \langle \prod_{i=1}^n V_i(p_i;z_i,\bar{z}_i)
        \prod_{j=1}^{b}W(\zeta_j,\bar{\zeta}_j)\rangle \;.
\ee
So any amplitude with $b$ boundaries corresponding to an array of D-branes
in imaginary time is identical to an amplitude without boundaries, but with $b$
additional insertions of the physical closed string state $|W\rangle$.
The $1/b!$ combinatoric factor ensures that the amplitudes exponentiate
properly to give a macroscopic closed string background.

\section{Conclusions}

We have generalized the prescription for computing closed string
amplitudes in the background of an array of D-branes in imaginary time
to arbitrary order, and used it to show that an amplitude with $b$
boundaries is identical to an amplitude without boundaries and with $b$
additional insertions of a particular physical closed string state $|W\rangle$.
Summing over the boundaries then corresponds to an insertion of
$\exp\int d^2z W(z,\bar{z})$, which corresponds to a closed string
background.

\section*{Acknowledgments}
We thank S.~Hirano, N.~Itzhaki and G.~Lifschytz for useful discussions.
This work is supported in part by the
Israel Science Foundation under grant no.~101/01-1.

\appendix

\section{A pair of branes in imaginary time}\label{app1}

The simplest configuration of branes in imaginary time which corresponds to
a {\em real} closed string background consists of two D-branes located
at $x_0=\pm ia$ \cite{a:GIR}.
The line of argument is similar to the one used in the array case.
We start with the amplitude for a pair of D-branes at $x=\pm a$ in real
space\footnote{We will suppress the external momenta and indices in this section.}
\be
 \widetilde{S}(P) = {1\over b!} \int \prod_{j=1}^b dk_j \,
 (e^{iak_j}+e^{-iak_j}) \widetilde{A}(k_1,\ldots,k_b)
 \delta\left(\sum_{j=1}^b k_j-P\right)\;.
\ee
Let us concentrate on the contribution of the brane at $x=a$,
{\em i.e.} the term with $e^{iak_j}$.
The contribution of the other brane will be similar, and we will
add it in later. We can assume that $a>0$ without loss of generality.
Fourier transforming the integrand with respect to $k_1$ gives
\be
\widetilde G(x,k_2,\ldots)=\int_{-\infty}^{\infty} dk_1\, e^{ik_1x}
\widetilde{A}(k_1,\ldots,k_b)
\prod_{j=1}^b e^{iak_j}\;.
\ee
The $k_1$ integration contour can be closed by an infinite semi-circle from above for
$x+a>0$, and from below for $x+a<0$ (Fig.~\ref{Contours2}).
\FIGURE[ht]{
\let\picnaturalsize=N
\def\picsize{3.1in}
\def\picfilename{ContoursSingleUp.eps}
\ifx\nopictures Y\else{\ifx\epsfloaded Y\else\input epsf \fi
\let\epsfloaded=Y
\centerline{\ifx\picnaturalsize N\epsfxsize \picsize\fi
\epsfbox{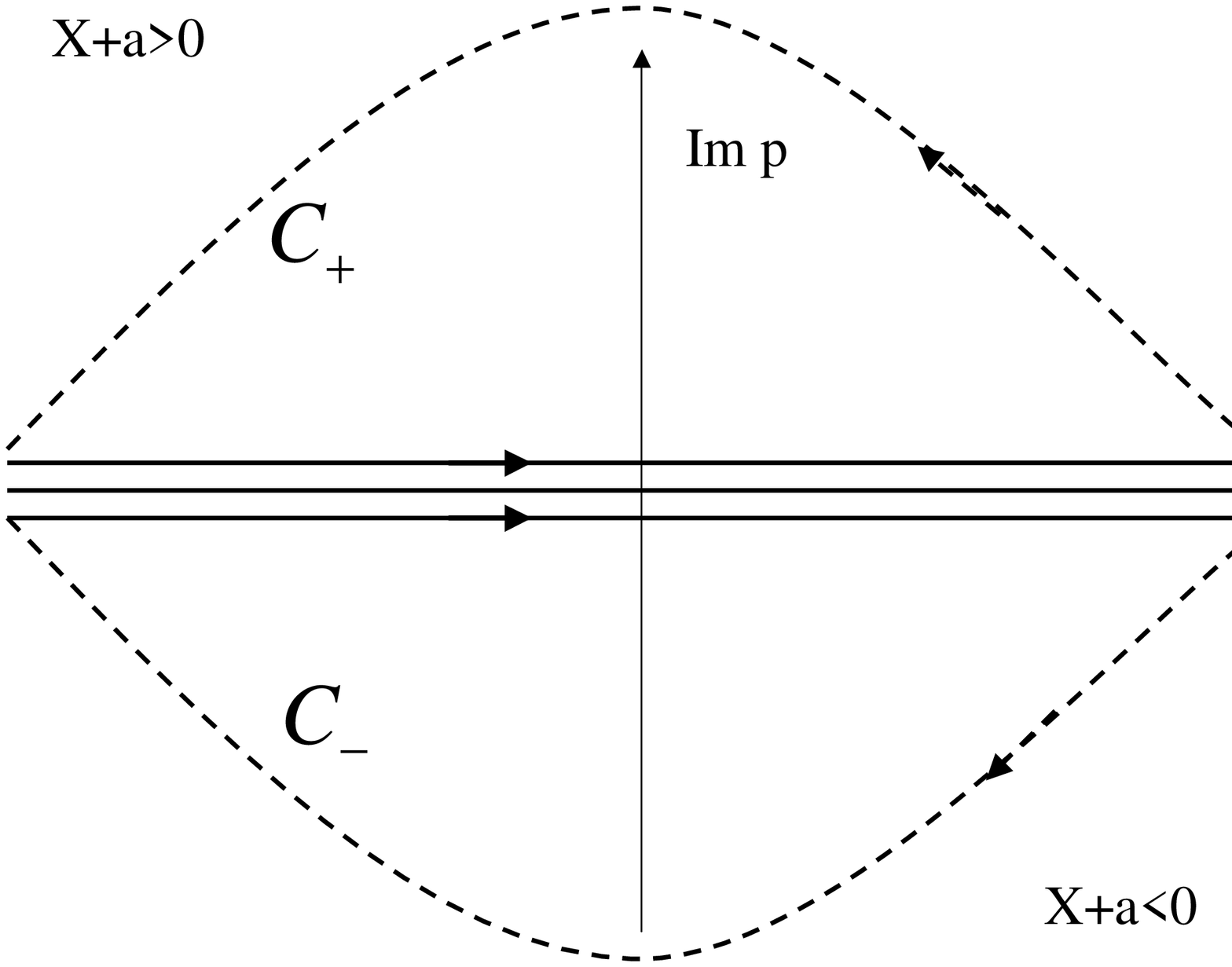}}}\fi
\caption{The two contours for $x+a>0$ and $x+a<0$}
\label{Contours2}
}
Assuming, as before, that the only singularities of $\widetilde{A}(k_1,\ldots)$
are on the imaginary $k_1$ axis, we deform the contour to ${\cal C}'_\pm$
(Fig.~\ref{ContoursC}).
\FIGURE[ht]{
\let\picnaturalsize=N
\def\picsize{3.1in}
\def\picfilename{ContoursSingleC.eps}
\ifx\nopictures Y\else{\ifx\epsfloaded Y\else\input epsf \fi
\let\epsfloaded=Y
\centerline{\ifx\picnaturalsize N\epsfxsize \picsize\fi
\epsfbox{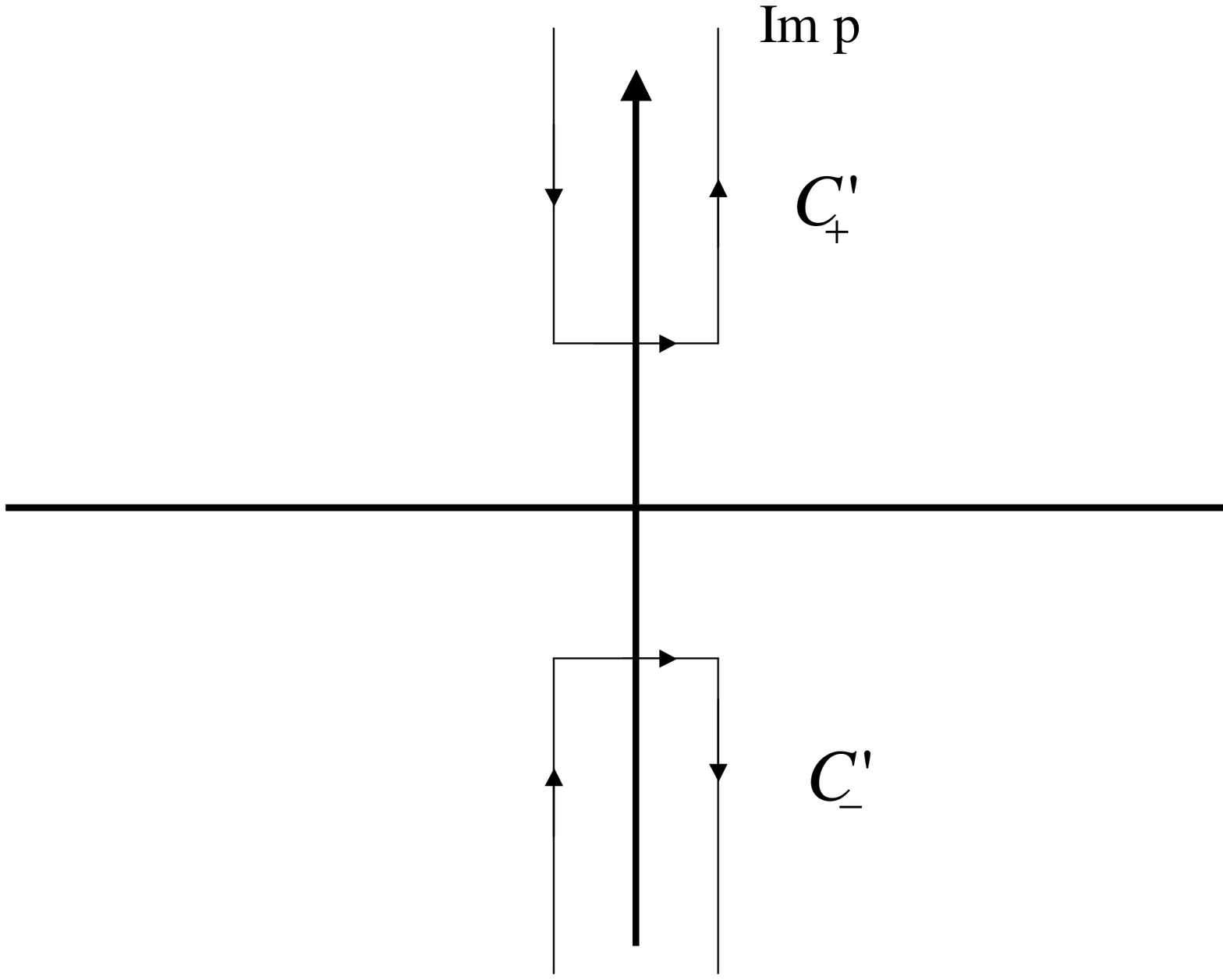}}}\fi
\caption{The two deformed contours for $x+a>0$ and $x+a<0$}
\label{ContoursC}
}
Back in momentum space we then get
\begin{eqnarray}
 \widetilde{S}(k_1,\ldots,k_b) &=& \int_{-\infty}^\infty dx\,
 \tilde G(x,k_2,\ldots)e^{-ik_1x}\nonumber\\
 &=& \int_{-\infty}^{-a}dx
\int_{{\cal C}'_-}dk\,
 e^{-i(k_1-k)x} e^{iak}  \prod_{j=2}^b e^{iak_j}
\widetilde{A}(k,k_2,\ldots,k_b)\nonumber\\
&+&\int^{\infty}_{-a}dx\int_{{\cal C}'_+}dk\,
 e^{-i(k_1-k)x}e^{iak} \prod_{j=2}^b e^{iak_j}
\widetilde{A}(k,k_2,\ldots,k_b) \;.
\label{split_contour}
\end{eqnarray}
Using\footnote{We define the $x$ integrals as
$\int_{a}^{\infty}dx\, e^{iwx}\equiv \lim_{n\to\infty}
 \int_{a}^{a+\frac{2\pi n}{w}}dx\, e^{iwx}$.
This vanishes unless $\omega=0$. Furthermore
$\int_{-\infty}^\infty d\omega \int_{a}^{\infty}dx\, e^{iwx} = \int_a^\infty dx\,\delta(x)$,
which vanishes for $a>0$ and is 1 for $a<0$.
This implies the above result.}
\be
 \int_{c}^{\infty}dx\, e^{iwx} = \left\{
\begin{array}{ll}
 \delta(\omega) & \;\; \mbox{for} \;\; c<0 \\
 0 & \;\; \mbox{for} \;\; c>0
\end{array}
\right. \;,
\ee
we obtain (after the Wick rotation)
\be
 S(E_1,\ldots,k_b) = \Theta(E_1) e^{-|aE_1|}
  \prod_{j=2}^b e^{iak_j}\,
  \mbox{Disc}_{E_1}\widetilde{A}(iE_1,k_2,\ldots,k_b)\;.
\ee
For the brane at $x=-a$ we replace $e^{iak_j}$ with $e^{-iak_j}$ in
(\ref{split_contour}), and exchange the contours ${\cal C}'_+$ and ${\cal C}'_-$.
The net result is to change $\Theta(E_1)$ to $-\Theta(-E_1)$ above.
Repeating the procedure for the other boundary momenta, and adding the
contribution of the other brane, gives
\be
 S(E) = {1\over b!}\int \prod_{j=1}^b dE_j\,
   S(E_1,\ldots,E_b)\, \delta\left(\sum_{k=1}^M E_i-E\right)\;,
\ee
where $E=-iP$ and
\be
 S(E_1,\ldots,E_b) = \prod_{j=1}^b \mbox{sign}(E_j) e^{-|aE_j|}\,
 \mbox{Disc}_{E_b}[...[\mbox{Disc}_{E_1}[\widetilde{A}(iE_1,\ldots,iE_b)]]...]\;.
\ee
For $b=1$ this reduces to the result in \cite{a:GIR}.
One can also reproduce the result for the periodic array (\ref{prescription})
by summing over an infinite number of pairs.

\end{document}